\documentstyle[aps,prl,epsfig,twocolumn]{revtex}
\newcommand{\beeq}{\begin{equation}}
\newcommand{\beeqar}{\begin{eqnarray}}
\newcommand{\eneq}{\end{equation}}
\newcommand{\eneqar}{\end{eqnarray}}
\begin{document}
\twocolumn[\hsize\textwidth\columnwidth\hsize\csname
  @twocolumnfalse\endcsname
\preprint{IMSc 96/}
\title{Quantum Chromodynamics and the Z- width }
\author{N.D. Hari Dass\dag} 
\address{The Institute of Mathematical Sciences, Chennai - 600 113, INDIA}
\author{V. Soni \ddag} 
\address{National Physical Laboratory, K.S. Krishnan Marg,New Delhi, INDIA}
\maketitle
\begin{abstract}

We show by explicit construction of an alternative theory for
   the strong interactions that it cannot be distinguished from
      QCD by any of the usual precision tests except on extending
	 these theories to the full electroweak theory, by the Z
	 paticle width.
\end{abstract}
\vskip 2pc] 

  Quantum Chromodynamics ( QCD ) is today the best accepted theory of the
  strong interactions. This is based on the fact that,

  1) Chiral Symmetry , which is the symmetry of the the hadronic
  interactions , and realised in a spontaneously broken phase with the
  pions as the the goldstone bosons, is also a symmetry of QCD.

  2) QCD is asymptoticaly free(AF) : this translates experimentally into the
  phenomenon of scaling in deep inelastic scattering(DIS).So far 
  DIS data has been consistent to good accuracy with the predictions of
  QCD.Even next to leading terms of QCD perturbation theory have been
  found necessary to explain certain high energy experiments.
  This is doubtless a very impressive record , hence the belief in QCD.
 While the relationship between AF and scaling is not straightforward, as
  argued by Gross,non-AF theories are unlikely to lead to scaling.

	  Furthermore, following the discovery AF it was
	  shown by Coleman and Gross \cite{gross} that nonabelian gauge fields are an
	  essential requirement for AF in four dimensions. 
	  
	  As far as the situation with theories with fundamental scalars
	  was concerned,depending on the choice of the scalar representation
	  AF could be obtained by fine-tuning the scalar self-coupling , but did not occur
	  naturally as in QCD. 

	  It should however be pointed out that  certain class of events
	  have recently been reported
	  which are not easily accommodated in QCD. These are , for
	  example, the rapidity
	  gap events seen at HERA \cite{hera} for  e-p scattering and the excess of
	  four jet events reported by the ALEPH collaboration. 
	  These suggest that perhaps a broader theory with gross
	  features of QCD is needed.

	  In what follows we propose such a broader theory in  which
	  QCD is extended by a chiral multiplet of scalar color- singlet fields that
	  interact only with the quarks.

	  We show in the next section that such a theory is not only
	  chirally symmetric but also AF in all the
	  couplings \cite {pre}. It is then important to see how
	  this theory fares versus QCD experimentally . 
          Our analysis for this theory shows that for deep inelastic scattering 
          the leading
          behaviour of this theory in the ultra violet is identical to that of
          QCD. This is so as  the Yukawa coupling and the
          scalar self coupling go to zero faster than the QCD coupling.In fact,
          for the case of 5 flavours, the numerical factors that arise are such
          as to make even the subleading QCD corrections to be more dominant than
          the contributions of these extra couplings.This is important as some
          high energy experiments are already sensitive to these subleading
          corrections.

	 The version of the theory chosen for this analysis is one in which
	 chiral symmetry is spontaneously broken or manifest with very small 
	 mass for the chiral multiplet. Such a theory
	 will then have scalar color singlet parton jets.Because of the
	 colour neutrality of these partons, on the one hand the multiplicity 
	 of the jets is expected to be low and on the other hand, large
	 rapidity gaps are expected between them . This would indeed be a very important
	 signature for this version of the theory.

	 We find that the usual strong interaction tests for QCD based 
          on the properties of quarks and gluons  will all go
	 through for this theory.
	 The theory cosidered here has , besides ,color singlet scalars,
	 but their couplings to the quarks are small in comparision to the 
	 QCD coupling ( see next section ). 
	 The other set of tests for the particle content of QCD 
	 include the
	 precision measurements like the R parameter and the  g-2 for the
	 muon.Here we will have extra contributions from our new scalar color
	singlet partons which couple to the photon. However , we find both
	these precision tests cannot confidently rule out our theory in favour of QCD.
	Finally, it is 
	the Z- width($\Gamma_Z$) ,
	perhaps the most precise number in
	particle physics,  
	that selects between the two theories.
\subsection{ The Theory}

The lagrangian is
\begin{eqnarray}
{\cal L} = & - & \frac{1}{2} (\partial_\mu \tilde\sigma)^2 -\frac{1}{2} 
(\partial_\mu \vec{\tilde\pi})^2
-\lambda^2 {(\tilde\sigma^2+\vec{\tilde\pi}^2-f_{\tilde\pi}^2)}^2\nonumber \\
& - & \overline\Psi_q\left[{\cal D}_\mu+
g_y(\tilde\sigma+i\gamma_5 \vec\tau\cdot\vec{\tilde\pi})\right ]\Psi_q
-\frac{1}{4}G_{\mu\nu}^a G^{\mu\nu a}
\label{lagan}
\end{eqnarray}
where ${\cal D}_\mu=\partial_\mu -ig_3A_\mu^aT^a $ and $G_{\mu\nu}^a = \partial_
\mu A_\nu^a - \partial_\nu A_\mu^a+ g_3 f_{abc} A_\mu^b A_\nu^c$. 
$A_\mu^a$ is the gluon field and $T^a$ the SU(3) generator in the fundamental 
representation.
$g_y$, $g_3$ and  $\lambda$ are the Yukawa, QCD, and scalar self- couplings 
respectively.
 $\overline\Psi_q$ is the quark field and $(\tilde\sigma,\vec{\tilde\pi})$
are the additional chiral scalar fields. Note that $\vec{\tilde\pi}$ should
not be confused with the traditional pions($\vec\pi$) with mass $\simeq$ 140 Mev, 
though it has the same quantum numbers.
The $\beta$ function for the QCD coupling, $\alpha_{s}$, is
\begin{equation}
\frac{\partial \alpha_{s}}{\partial  t} = - \left ( \frac{33 - 2 N_F}{3}\right )
 \frac{\alpha_{s}^2}{8\pi^2} \hspace{5em} \left ( g_3^2=\alpha_{s} \right )
\label{dadt}
\end{equation}
This is for $m_q=0$ and $t=\ln (p/\mu)$.\\
To 
one-loop order the $\beta$-function
for the QCD coupling does not receive any contribution from the Yukawa coupling, $g_y$, or the scalar self
coupling $\lambda$, as the chiral multiplet is colour singlet.

The Yukawa coupling $g_y$ for the pion and sigma to the quarks has the following $\beta$
function (assuming 3 colours) 
\begin{equation}
\frac{\partial {g_y^2}}{\partial t} = \frac{g_y^2}{8\pi^2}
\left [ 12 N_g^` g_y^2 - 8 \alpha_{s} \right ]
\label{dgy}
\end{equation}
where $N_g^`$ is the number of generations to which the chiral multiplet 
couples.\par

Following \cite{schre} we can now define the ratio $\rho=g_y^2/\alpha_{s}$ 
and write the following equation
for $\rho$ using Eqs~(\ref{dadt}) and (\ref{dgy}).
\begin{equation}
\frac{\partial \rho}{\partial {\alpha_{s}}} = -\frac{\rho}{\alpha_{s} A}
[12 N_g^` \rho -8 +A]
\label{DrByDa}
\end{equation}
where $A=(33-2N_F)/3$ and $N_F$ is the number of flavours that effectively
couple
to the gluons. For the $N_F=6$ case,
 there are two regimes.Calling $\rho_c = 1/12$, we
 have\\
 {\bf The Region $0<N_g^` \rho<\rho_c$}: 
 
 In this case $\partial \rho/\partial \alpha_{s}>0$. This implies that 
$\rho$ decreases as $\alpha_{s}$  decreases, that is, $\rho$ will decrease 
with increasing momentum scale. We can integrate the $\rho$ equation to get
\begin{equation}
\rho=\frac{\alpha_{s}^{1/7} K}{(1+12N_g^` \alpha_{s}^{1/7} K)}
\end{equation}
where $K$ is a positive integration constant that is set by initial data on 
$\alpha_{s}$ and $g_y^2$. It is therefore clear that there is a whole family of
solutions corresponding to different K's. 
Deep in the ultraviolet when $\alpha_{s}\rightarrow 0$ 
$$N_g^` \rho\sim K\alpha_{s}^{1/7}$$
 which further implies that in the ultra violet 
$$N_g^` g_y^2 \sim K\alpha_{s}^{8/7}$$
 This means that $g_y^2$ is asymptotically free and vanishes faster than 
 $\alpha_{s}$. Therefore, the leading behaviour of this theory in 
the ultraviolet is
 given by the QCD coupling with the Yukawa coupling contributing only in 
sub leading order. \par
The region $N_g^` \rho > 1/12 $ is also of interest, though from a different
physical perspective. However, the theory is not AF in this region. Therefore
 we shall not consider it anymore in this paper. See however, refs \cite{pre} 
and \cite{umb}
for further discussions of this region.\par

The above analysis is valid for $q^2 \ge m_t^2$. For the region
$m_b \le q \le m_t$, the relevant behaviours are : $\rho_c = 1/36,
N_g^\prime \rho \simeq K\alpha_{s} ^{1/23}$. 
For $N_F \le
4$. 
there is no AF regime . 
The solution for $N_F = 4$ is $\rho /(\rho + 1/36) = K \alpha_{s}
^{-1/25}$.

For the purposes of our paper (in the context of the discussion on
$g - 2$ for muons) it is important that regions where QCD is
perturbatively treatable, our theory is too. 
AF for the scale $ > m_b $ implies $\rho \le 1/36$.
Thus at $q = m_b$,the Yukawa couplings are small. Now the lack of AF for
$q < m_b$ has the desired effect of making $\rho$ even smaller as we go
to
smaller energy scales. Thus $g_y,\lambda$ are perturbatively
treatable wherever QCD is. 

The $\beta$-function for $\lambda$ in our model is 
\begin{equation}
\frac{\partial \lambda}{\partial t}=\frac{1}{8\pi^2}\left[
2\lambda^2-144 N_g^` g_y^4+24 N_g^` g_y^2\lambda\right]
\end{equation}
By defining the ratio $R=\lambda/g_y^2$\cite{schre} we can convert to an equation
for $R$ that depends on the single variable $\rho$ and find that
\begin{equation}
\frac{\partial R}{\partial \rho}=\frac{1}{[12\rho-8+A]}
\left[ 2R^2+R \left(12+\frac{8}{\rho}\right)-144 \right ]
\end{equation}
For related issues in the standard model see \cite{ross,Hil,Zim,KSZ,tana,KTY,
KMZ,har}.

It is found that only on a single trajectory in the $[R, \rho]$ parameter 
space, that is the so called invariant
line~\cite{schre,pre,umb} behaviour of $R$ for the regime
$N_g^` \rho <1/12$ is such that $R\rightarrow 0$ in the ultraviolet. 
It follows further that $\lambda\rightarrow 0$ in the extreme ultraviolet 
even faster than $g_y^2$. 

Thus we have classes of theories that are not only AF in all their couplings,
 but become increasingly indistinguishable from QCD at high energies. 
As far as AF is concerned 
one loop analysis is stable against higher loop corrections 
\cite{gross}. 
Since these classes of theories are AF, they are all candidates for a consistent theory of strong interactions.

\subsection {	The R parameter}
	
	The R parameter measures  the ratio of 
		  $ \sigma ( e^+e^- \rightarrow hadrons )$ to $ \sigma (e^+e^-
		  \rightarrow \mu^+\mu^-)$. 
In QCD , at high energies, the former is approximated by 
	    $ \sigma( e^+e^- \rightarrow \Sigma q\bar q )$.
When the
energies are well below the Z mass we can neglect the contribution
of the Z mediated process to R.To leading order ,then, the R parameter
measures the total number of operational flavours
(of course, multiplied by the number of colors of the quarks ), 
modulated by their charge squared .

As we move to
higher generations the  R parameter changes rapidly at quark mass
thresholds.At thresholds we also encounter many resonances which also
give rapid changes in the R parameter. However, away from thresholds R
can be quite stable. Therefore in the region above $b\bar b$
threshold we expect R to be stable apart from QCD corrections.But as we
approach ,   $\sqrt s  = M_z$,a new class of diagrams become operative and the R
parameter has a steady rise to the Z peak.

 There is thus an energy region  ,20- 40Gev , where the effect
 of the Z is yet very small , where R is relatively stable. To leading order
 , that is , in the absence of QCD corrections,here
				$R_0 =  11/3$.

The R parameter will change for our theory as we have new scalar charged
partons , the $\tilde\pi^+\tilde\pi^-$ that couple to the photon which will
contribute to the hadronic cross section. For the the zero mass partons
considered by us the contribution to the R parameter is exactly
calculable and is given by
		  $R = 1/4$.
This additional contribution should be clearly visible, particularly in the region:
20 - 40 Gev .The contribution of Z - exchanges are negligible in this
region.

The R parameter receives QCD corrections and  
the QCD corrected R parameter in this region is:
               $R(s) = R_0 ( 1 +  \alpha_{s} /\pi +...       )$
The measured value of the R parameter in this region has a world average of
4.02. The difference between this number , 4.02 , and  $R_0 = 11/3$ is supposed
to come from the QCD corrections and yield a value for $\alpha_{s}$ 
at  this energy scale.

However if we add the extra contribution of our new partons,
		$R_0  = 11/3 + 1/4  = 3.91$,
leaving only a deficit of .11 to be accounted for by the QCD corrections.The
corresponding value of the QCD coupling will then be too low to be admissable.
However  presently the systematic errors at Amy, Topaz etc are not so small , of the order of
5\% . Also, different groups have reported  R in this region to be as high as 
4.2 or even as low as 3.8 . This circumstance means that our theory cannot be ruled out
as the effect we are considering is 
				 $\Delta R/ R  =  6$ \%.
It is worth pointing out that experimentally low multiplicity ($\simeq
5$) jets are not counted as these are very unlikely in QCD. On the
other hand, for the color neutral pionic partons of our theory, we
expect to have only 
low multiplicity jets most of which would have been
excluded by experimental cuts. Of course, two prong events would have
been counted as $\mu^+\mu^-$-pairs and could have shown up as
anomalies which could be distinguished by their different angular 
dependence 
as compared to  leptons.

Given these facts the R parameter is not at present a definitve test
for this theory versus QCD  though reduced systematic errors could 
put things on the borderline.

\subsection{g-2 for the muon}

 The contribution to g-2 ,the muon magnetic moment, can be potentially disturbed
 by the presence of additional charged particles. 
 Our theory has particles with zero or light
 masses coupling to photons thereby contributing additionally to the photon
propagator. Such
 a contribution to the propagator can be related to the contribution to the
 R parameter we have just considered . This is precisely how g-2
 is calculated in the literature \cite{ynd}. In this work
 the contribution to g-2 for the muon , $a_v$ , is given by
 \beeq
                   a_v =\int_{4m^2} dt K(t)R(t)
 \eneq
where $R(t)$ is the R-ratio and $K(t)$ is a known function.

  However , the manner in which zero mass scalar partons enter the low energy description 
  of R(t) ( where $\sqrt t$ is the centre of mass energy ) is subtle. At low energy there is  
  mixing 
  between our zero mass partonic pions and the pseudoscalar channel . 
  This will generate a higher mass state with quantum numbers of the
  pion in addition to the the usual 'goldstone pion ' associated with the spontaneous breaking of
  chiral symmetry. 
  Since we cannot calculate the mass of this non-perturbatively
  generated extra state 
  we can only use experiment to glean its mass. The particle data book lists the first 
  additional state with pion quantum numbers at 1.3 Gev. At low energies then
  we must use this state  to calculate the extra contribution to
  the photon self energy or the R parameter as the usual pion's
  contribution is already accounted for in the standard treatment of
  hadronic contributions to g-2.The threshold for the contribution of this state
 then starts at $\simeq $2.6 Gev. This puts us in the perturbative QCD regime. 

  Before computing the extra contributions a few remarks are in order: 
1) The low energy regime,0.8-2.0 Gev, has to be gleaned from experiment as perturbative 
QCD can not be used .The
contribution to $a_v$ from this region  as listed in  Table 2  of \cite{ynd} is
	    $(1404 \pm 100 ) . 10^ {-11}$
As this is evaluated from experimental data, one cannot differentiate the 
contributions from QCD and our theory .
2) Perturbative QCD is used for $t >2 Gev^2$ ( Table 1 of
\cite{ynd}) except
for the threshold regions  which are populated by numerous 
resonances,
where again one has to only rely on experimental data, 
and can not differentiate between our theory and QCD.
These regions are:
3.3- 3.6 Gev,3.6 - 4.9 Gev and  9 - 14 Gev.

  Thus it is only for regions for which estimates  are made via
  perturbative QCD that comparison between this theory and QCD is possible.
In respect 
  of the foregoing discussion this region for us must begin at  $\sqrt t >2.6$ 
  Gev. 
  We briefly sketch how the additional contribution can be evaluated 
for our theory in the 
  region  2.6 - 3.1 Gev:
  i) K(t)  goes as  $1/t^2$. Assuming $R(t) =  R_0$ and using eqn(10)
 we can get the two contributions   for 1.4-2.6 Gev and  2.6-3.1 Gev
for QCD.
The partial contribution for the region ,2.6 -3.1 Gev is found to be 
1/10 of the total.
ii) The extra contribution to R assuming zero mass pionic partons is  
$\delta R$ = 1/4 whereas the
QCD contribution is  $R_0  = 2$.
Thus the fractional extra contribution is  1/8.
iii) This is further down when we take into account the mass of the excited 
state (1.3 Gev). A rough 
estimate is provided by mutiplying by the phase space factor 
$( 1  -   4 M ^ 2/t ) $  =  0.17
taken at the average value 2.85 Gev for $\sqrt t$.  
The total extra contribution  for the region 2.6-3.1 Gev works out to
 $1.2 \cdot 10^{-11}$. Below we display the contributions and errors 
 for various regions:
\begin{tabular}{lllll}

$~~~\sqrt t~~~$ &  ~~~    QCD~~~   &~~~         Chiral~~~    &~~~         Theo.~~~   &      ~~~    Sys.~~~\\
in Gev	       &	      & multiplet                         &           error    &     error($ 5\%$ )\\

 2.6 - 3.1   & ~~~       56              &         1.2                   &       $\pm 1$        &   $ \pm$    2.8\\

 4.9 -  9     &  ~~~      67.5            &         4                     &       $ \pm 1$        &   $\pm$    3.5\\

   $>$  14       & ~~~       13              &         0.9                   &       $\pm 2$        &   $  \pm$   0.65\\

\end{tabular}

   It should be noted that:
   i) The theoretical error in \cite{ynd} is arbitrarily estimated as 
   half the $\alpha_{s}^2$
   correction to R.
   ii) we have taken the systematic error to be  5\% of R (see Sec 3)

 The extra contribution of our theory  falls within the sum of 
   the theoretical and the systematic errors .
 The error in the low energy region , 
   0.8-2 Gev is roughly $100\cdot  10^{-11}$. By comparison, all our extra 
contributions are negligible.

   We are therefore led to the conclusion that  g-2 for the muon despite being a very high
   precision measurement of the charged-particle content of theories 
cannot differentiate between QCD and our 
   theory  -  a rather non trivial result.

\subsection{   Z width}

The Z-width data on the other hand is known with great accuracy.
The minimal coupling of the chiral multiplet to $Z_{\mu}$ 
is \cite{long,aleph}:
\beeq
{\cal L}_{lin}^{neut}= e(A_{\mu}-{\gamma\over 2} Z_{\mu})
(\vec{\tilde\pi}\times
\partial_\mu\vec{\tilde\pi})_3
-{e\over 2cs}Z_\mu(\tilde\pi_0\partial_\mu\tilde\sigma-\tilde\sigma
\partial_\mu\tilde\pi_0)
\eneq
where $\gamma = (1-2s^2)/cs)$ with $s$ being $sin \theta_W$ and $c^2 = 1-s^2$. 
The contribution to the hadronic width of Z-boson due to the extra scalars can be calculated easily:
\beeq
{\Delta\Gamma^Z\over \Gamma^Z_{had}} = {9((1-2sin^2 \theta)^2+1)\over N_c(90-168sin^2\theta + 176 sin^4\theta)}
\eneq
At $sin^2 \theta \simeq .25 $,this works out to roughly 4.5\% of the total width $\Gamma^Z$.The high precision LEP data only allows total uncertainty of about
.3\%.\par 
This more or less immediately rules out the extended version where the
chiral symmetry in the extended sector is spontaneously broken or where the
chiral symmetry in the extended sector is manifest with low mass for the
multiplet.

  A different version of this theory  where the chiral multiplet mass
  of over one half the  Z width will be exempt from this problem 
  This will be considered separately.\cite{long,aleph}
\subsection{   CONCLUSION}

   By explicit construction of an alternative theory to QCD for the strong interactions we
   have found that all precision tests for QCD except forthe Z width cannot select between the 
   two theories. Only by extending the theories to the FULL electrweak standard model do we find
   an unambiguous support in favour of QCD from the Z width.
   This underscores the fact that most tests and vindications of QCD that are to be found in 
   archival refrences in the literature are just not adequate. From  this work we find that 
   only the Z width is the final arbiter.

\end{document}